# Macroscopic polarization in the nominally ergodic relaxor state of lead magnesium niobate

Lukas M. Riemer[1,a)], Kanghyun Chu[1], Yang Li[2], Hana Uršič[3], Andrew J. Bell[2], Brahim Dkhil[4], and Dragan Damjanovic[1]

[1]Group for Ferroelectrics and Functional Oxides, Swiss Federal Institute of Technology in Lausanne – EPFL, 1015 Lausanne, Switzerland
[2]School of Chemical and Process Engineering, University of Leeds, Leeds, LS2 9JT, U.K.
[3]Electronic Ceramics Department, Jožef Stefan Institute, Jamova Cesta 39, 1000, Ljubljana, Slovenia
[4]Laboratoire Structures, Propriétés et Modélisation des Solides, Centrale Supélec/CNRS UMR8580, Université Paris-Saclay, F-91192 Gif-sur-Yvette, France

[a)]**Author to whom correspondence should be addressed:** lukas.riemer@epfl.ch

**Abstract:**
Macroscopic polarity and its dynamic response to external electric fields and temperature in the nominally ergodic relaxor phase of pristine lead magnesium niobate crystals and ceramics, $Pb(Mg_{1/3}Nb_{2/3})O_3$ (PMN), were investigated. Dynamic pyroelectric measurements provide evidence for persistent macroscopic polarity of the samples. Annealing experiments below and above Burns temperature of polarized samples relate this polarity to the presence of polar nano entities and their dynamics. The dc electric field strength required for macroscopic polarization reversal is similar to the amplitude of the ac field where dynamic nonlinear dielectric permittivity reaches maximum. Consequently, the aforementioned maximum is related to the reorientation of polar nano entities. The results question the existence of an ergodic state in PMN below Burns temperature.

Relaxor ferroelectric solid solutions are at the forefront of piezoelectric technology and focus of continued research activity.[1,2,3] Despite sixty years[4] of ever-growing experimental and theoretical research a predictive model capable to describe the majority of available experimental facts is still missing. In fact, the opposite is true. The ground state of relaxor ferroelectrics and their relaxor endmembers still spark controversial discussions in the field of ferroelectrics and beyond.[5,6,7,8,9,10] In particular, the correlation and dynamics of macro-, meso-, and nanoscopic-scale polarization and its impact on piezoelectric properties remains an enigma.[3,11,12,13]

Common characteristics of relaxors are a broad maximum in the dielectric permittivity measured as a function of temperature and dielectric dispersion below the temperature of maximum permittivity ($T_m$).[1,2] Most theoretical models correlate these characteristics to relaxation of polar dipoles.[1,6,14,15,16] Temperature regions of relaxor systems can be classified based on their macroscopic dynamic, structural and dielectric properties as paraelectric, ergodic relaxor, non-ergodic relaxor, or ferroelectric.[2,17] Note that the terms paraelectric and ferroelectric imply ergodic and non-ergodic behavior respectively. In the mesoscopic description, the coexistence of static and dynamic nano-sized regions of correlated polarization is commonly accepted.[18,19] However, there is no consensus in the literature about the geometric interpretation or the terminology used to describe aforementioned regions.[20,21] To emphasize this unresolved debate the generic term "polar nano entities" is used in this manuscript, even when quoting references that use different terminologies.

Several characteristic temperatures were introduced to describe different states of relaxors: the Burns temperature, $T_B$, the coherence temperature, $T^*$, and the freezing temperature, $T_f$.[18,19,20] At $T_B$ a transition from the paraelectric state to the ergodic relaxor state takes place. This transition is explained as the first appearance of dynamic polar nano entities.[18,19] At $T^*$ static regions of correlated polarization emerge.[18,19] Below $T_f$ the system is said to be in the non-ergodic relaxor state in which the polar structure is frozen[22] in an out-of-equilibrium state that depends on the thermal and field history, as opposed to the ergodic state. While the onset of non-ergodic behavior in the vicinity and below $T_f$ was investigated intensively in the early 1990s,[22,23,24] PMN was still referred to as paraelectric at temperatures above $T_f$.[25,26] Later on, the term ergodic relaxor state was introduced to distinguish the temperature region between $T_f$ and $T_B$ from the paraelectric state above $T_B$.[2,27]

$Pb(Mg_{1/3}Nb_{2/3})O_3$ (PMN) is presumably the best-studied ferroelectric relaxor to date. Until now, its ergodic relaxor state (above $T_f \sim 217$ K)[22] is usually described as macroscopically cubic and centrosymmetric with locally broken symmetry within polar nano entities.[2,28,29] In contrast, recent studies find macroscopic piezoelectric response between 300 K and 770 K.[30] Apparently, neither the ground state of the ergodic phase nor its dynamic response to external electric and mechanical fields is fully understood. Dielectric tunability and the reorientation of polar nano entities have been considered viable mechanisms to explain the specific nonlinear dielectric response of PMN, but some inconsistencies in proposed models and experimental results are still unresolved.[11,15,31,32] To clarify both the nature of the ergodic state and polarization dynamics under external electric field, we investigate zero-bias and electric field-induced macroscopic polarity of PMN in the nominally ergodic relaxor state and its dynamic behavior combining pyroelectric current, and nonlinear dielectric permittivity measurements.

Experiments were carried out on a batch of PMN ceramics and crystals of two sources grown with different methods. Two plates of high quality (001) cut PMN single crystals denoted as PMN-A and PMN-B with dimensions of (2.2 x 2.3 x 1.0) mm³ and (2.8 x 2.0 x 0.7) mm³ were prepared. Crystal growth conditions are described in the supplementary material. Preparation of the ceramics is described elsewhere.[33] All samples were polished and sputter coated with gold or platinum. The methods for nonlinear dielectric permittivity[11] and pyroelectric current[34] measurements are described elsewhere. Pyroelectric current measurements were performed with a temperature rate of ± 0.075 K/s. All pyroelectric and dielectric experiments were performed at ambient temperature.

Measurements of dynamic pyroelectric current are a convenient, non-destructive way to verify whether materials possess macroscopic polarization $P_M$. When measured under dynamic thermal conditions pyroelectric currents $i$ result dominantly from the temperature dependence of $P_M$:

$$i = A \frac{\partial P_M}{\partial T}\frac{\partial T}{\partial t} = A \times p \times \frac{\partial T}{\partial t}, \qquad (1)$$

where $A$ is the sample area, $T$ is the temperature, $t$ is the time, and p is the pyroelectric coefficient.[34] According to Equation 1 a temperature modulation with a triangular waveform of a polar material is expected to result in a square wave current response,[34] if $P_M$ changes sufficiently quickly with temperature. Moreover, a direct correlation between the sign of the pyroelectric response and the orientation of $P_M$ is given, i.e. the sign of the pyroelectric current obtained for a positive temperature rate changes if the spontaneous polarization vector is reversed (e.g., if the sample is flipped over). In the case of a non-polar dielectric material, a change in temperature does not result in a pyroelectric current, unless the sample is biased by a dc electric field during the measurement.

Pyroelectric current measurements of pristine PMN-A crystal performed at temperatures around 295 K are presented in Figure 1a. The crystal had not been subjected to an electric field before those measurements. At this temperature, approximately 75 K above $T_f$ and 25 K above $T_m$, the material is supposed to be in the nominally macroscopically non-polar ergodic relaxor state. Contrary to expectations, a clear pyroelectric square wave response is evident. To exclude artificial pyroelectric response that might arise in case of a temperature gradient across the sample[34] or burden voltage of the measurement device[35] the sample was flipped over and measured again. As expected for a material with persistent macroscopic polarization, the pyroelectric response changes its phase by 180° when the sample is flipped over. Similar experiments were performed for crystal PMN-B and ceramic samples in pristine and annealed conditions, respectively (Figure S1). In all cases periodically modulated pyroelectric response was measured. This is a direct proof of persistent macroscopic polarity in the ergodic phase of pristine PMN.

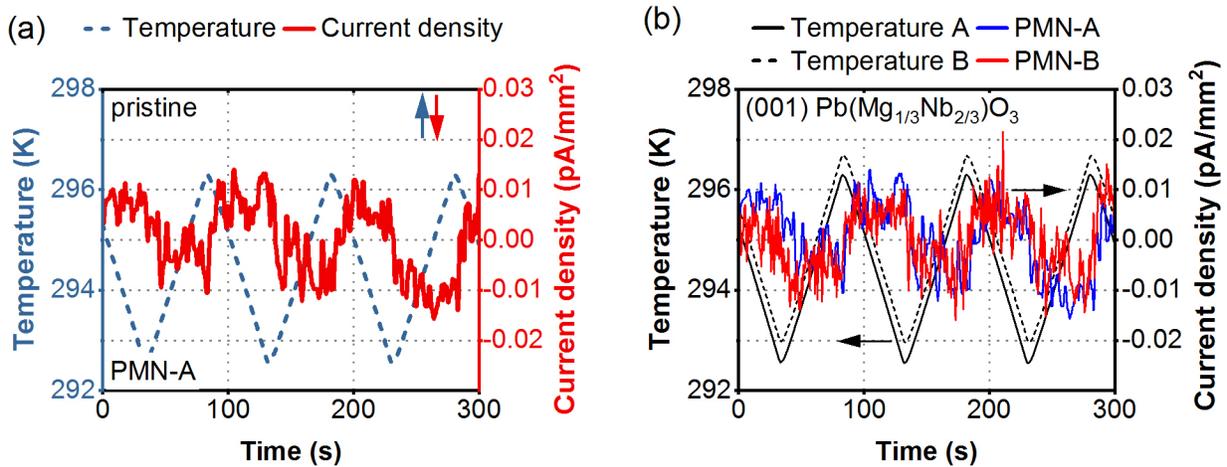

Figure 1: Proof of persistent macroscopic polarization in the ergodic phase of ferroelectric relaxor lead magnesium niobate by pyroelectric measurements of pristine (001) PMN crystals of different sources. Qualitatively similar response is obtained after annealing at 773 K, for samples with Au and Pt electrodes, and for ceramic samples. (a) A triangular temperature modulation of crystal PMN-A results in a square wave current response as expected for materials with persistent macroscopic polarization, Eq. 1. A pyroelectric coefficient of (0.073 ± 0.010) µC/m²K can be determined. A blue and a red arrow indicate the relation between the sign of temperature rate and the sign of current density. (b) Comparison between crystals PMN-A and PMN-B. The striking similarity obtained for crystals of different sources provides compelling evidence that the observed behavior is not sample specific. For crystal PMN-B data were shifted by 180 degree to be in phase with the data of crystal PMN-A.

Furthermore, the striking similarity obtained for crystals of different sources (Figure 1b) provides compelling evidence that the observed behavior is not sample specific. For PMN crystals pyroelectric current densities of about (0.005 ± 0.001) pA/mm$^2$ could be measured resulting in an average pyroelectric coefficient of (0.073 ± 0.010) µC/m$^2$K. For PMN ceramics pyroelectric coefficients of about (0.063 ± 0.027) µC/m$^2$K were obtained. The difference, which is within experimental error, might be explained by long-range symmetry-breaking strain fields of grain boundaries[36] or differences in homogeneity, stoichiometry or defect concentration, which to some degree are present in any material.[37] The calculated pyroelectric coefficients of PMN are approximately two orders of magnitude smaller than the pyroelectric coefficient of PVDF[38] and approximately four orders of magnitude smaller than the pyroelectric coefficients of (1-$x$)Pb(Mg$_{1/3}$Nb$_{2/3}$)O$_3$–$x$PbTiO$_3$ single crystals with morphotropic phase boundary composition.[39]

Pyroelectric and piezoelectric effects have been reported in ~500 nm thick PMN films in the past.[40] However, all thin films are inherently asymmetrical and the origin of the polarization cannot be unequivocally assigned to properties of the relaxor. For example, the pyroelectric coefficient was two orders of magnitude higher than in single crystals and ceramics examined in this study, suggesting contribution of thin film-related processes such as clamping from the substrate and electrode asymmetries.[41] Symmetry breaking in nominally centrosymmetric oxides has also been reported for amorphous strontium titanate films[42], barium titanate[43], lead zirconate[43], and barium strontium titanate[44]. Potential mechanisms involve inter alia, octahedral rotation[42], polar entities[43,44], defects[44], and strain gradients[44].

In agreement with current results, previous studies demonstrated piezoelectric response in the ergodic state of PMN using resonant piezoelectric spectroscopy (RPS) and resonant electrostriction spectroscopy (RES).[30] A breaking of macroscopic centrosymmetry was hypothesized to result from alignment of dynamic polar nano entities in stress gradients of chemical ordered regions.[30] In contrast to pyroelectric measurements, RPS and RES cannot prove macroscopic polarity inasmuch as electrostriction is a general property of dielectric solids[45,46] and piezoelectricity merely requires breaking of macroscopic centrosymmetry.[46,47]

Following the hypothesis of concerted alignment of polar nano entities, the direction of the macroscopic polarity revealed in our work should be temporarily changeable. To test this hypothesis, pyroelectric measurements were conducted before (Figure 1a) and after an electric dc field of 10 kV/cm was applied against the initial polarization of the pristine PMN-A sample (Figure 2a). When compared to the same crystal in pristine condition, two significant changes are seen after the application of the electric dc field against initial polarization. First, the relation between the sign of temperature rate and the sign of the pyroelectric current density changed implying the reversal of macroscopic polarization. The minimal dc field required for this reversal was determined to be $E_m$ = (1.35 ± 0.05) kV/cm both in crystals and ceramic samples. Second, the amplitude of the pyroelectric current density increased by more than order of magnitude with respect to the pristine sample. The temporal stability of the electric field induced state was studied by subsequent pyroelectric measurements as depicted in Figure S2. In contrast to the slower stretched exponential decay of dielectric permittivity observed below $T_f$[48], the decay of the pyroelectric response can be fit by Curie von Schweidler (CvS) law ($f(t) = f_0+(t/\tau)^n$)[49], where $f_0$ defines the offset, $t$ is time, $\tau$ is the time constant and $n$ the power exponent. A power exponent of n = -0.235 ± 0.001 was derived, in good agreement with previous reports, in which the decay of metastable surface piezoresponse was studied locally in PMN crystals.[49] The induced piezoelectric response was attributed to the realignment and subsequent relaxation of dynamic polar nano entities.[49] Relaxation mechanisms following CvS law were previously assigned to motion of phase boundaries of polar nano entities[50] which can become significant under large electric fields.[51]

To accelerate the relaxation of the field-induced state, the polarized PMN-A crystal was annealed at elevated temperatures and the pyroelectric response was measured after cooling to ambient temperature. An interesting fact highlighted by this experiment is that the direction of macroscopic polarization does not return to its initial, pristine state if the sample is annealed below $T_B$. The initial orientation is only recovered after annealing above $T_B$ as shown in Figure 2b-c. Such "memory effect" has been previously observed in relaxor ferroelectric solid solutions.[52] Thus, a correlation between the direction of macroscopic polarization and $T_B$ indicates that at least a part of the induced polarity is directly related to dynamics of polar nano entities. After both annealing temperatures the amplitude of the pyroelectric current density is around 0.006 pA/mm² but opposite of sign, as shown in Figure 2. It is therefore likely that the majority of the induced polarity decays during annealing while a smaller fraction is more persistent. Several reports have mentioned hierarchical relaxation processes with different activation energies.[37,50,53]

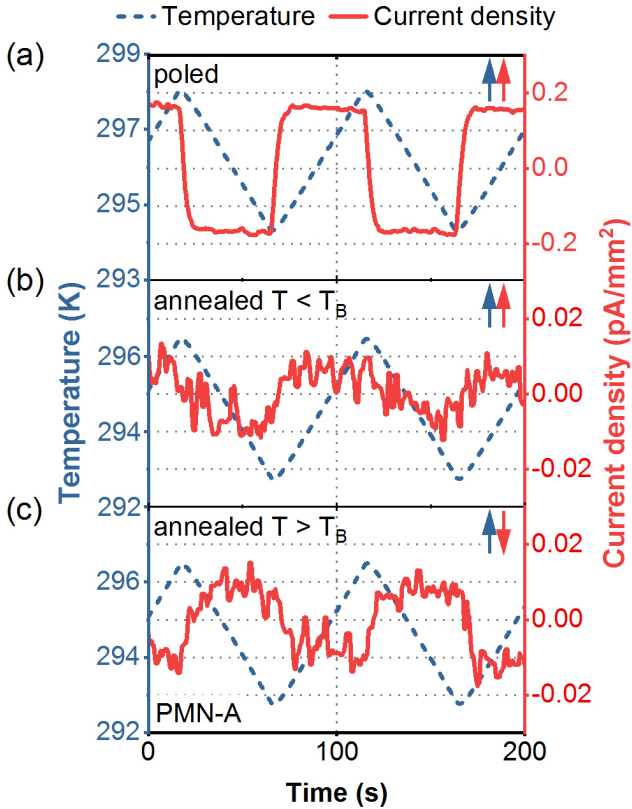

Figure 2: Pyroelectric current of poled and annealed (001) PMN-A crystal. A blue and a red arrow indicate the relation between the sign of temperature rate and the sign of current density. Poling against the initial direction of polarization results in persistent macroscopic polarization of opposite direction. A minimum field of $E_M=(1.35 \pm 0.05)$ kV/cm is required for this polarization reversal. Annealing above Burns temperature ($T_B \sim 630$ K)[19] restores the initial orientation of polarization. The ordinate of the pyroelectric current density in Fig. 2a is stretched by a factor of ten in relation to Fig. 1 and Fig. 2b-c. All measurements were performed with identical sample orientation. (a) Poled at 10 kV/cm, (b) poled at 10 kV/cm and annealed at 573 K, (c) poled at 10 kV/cm and annealed at 673 K.

To investigate the nature of macroscopic polarization in more detail the dynamic dielectric permittivity of PMN was measured as a function of driving field amplitude and frequency. It is assumed that the nonlinear response is sensitive to dynamics of polar nano entities.[11,31,54,55,56] As can be seen in Figure 3, neither the field nor the frequency dependence of the dynamic dielectric response between 1 Hz and 1 kHz itself allow for any further conclusion beyond the discussion in Ref.[11]. However, a striking similarity between the electric field needed for macroscopic polarization reversal, $E_M$, and the electric field amplitude of maximum nonlinear dielectric permittivity was found. Similar results were obtained for crystals and ceramics as summarized in Figure S3. In all samples the driving field amplitude of maximum dynamic dielectric permittivity depends weakly on sample history and measurement settings, e.g. the time taken at each driving field amplitude and the field amplitude increment, as presented in Figure S4. Nevertheless, the maximum dielectric permittivity observed in PMN crystal and ceramic samples can be estimated to occur around (1.50 ± 0.25) kV/cm. It was speculated previously that this maximum could either be attributed to intrinsic, lattice electric tunability or the flipping of polar nano entities.[11,31,32] Intrinsic electric tunability is a reversible process[57,58] that cannot explain reversal of persistent polarization. The fact that polarization reversal field $E_M$ coincides with the field at which the nonlinear dielectric permittivity reaches its maximum thus strongly suggests that the underlying mechanism is related to switching of polar nano entities and not to lattice tunability.

Furthermore, it can be shown that the fields used in this study do not induce a ferroelectric state in PMN samples. In principle, a ferroelectric phase can be induced in PMN by field-cooling or by application of bipolar fields under isothermal conditions.[17,59] The smallest threshold fields required to induce a ferroelectric phase by field cooling with dc bias applied along <111> or <100> directions are reported in the literature to be around 1.8 kV/cm and 2.9 kV/cm respectively.[17] At ambient temperature the threshold fields increase to more than 25 kV/cm.[59] For bipolar driving fields under isothermal conditions the threshold fields are comparable or greater.[17,59] The characteristic electric fields reported in this study (1.35 and 1.50 kV/cm) and the used poling field (10 kV/cm) are significantly smaller than that and thus cannot enable percolation of polar nano entities into a ferroelectric state at ambient temperature.

In summary, experimental evidence for macroscopic polarity in the nominally ergodic relaxor phase of pristine PMN is presented. The orientation of macroscopic polarization can be switched with the application of an electric field and requires subsequent annealing above Burns temperature to return to its initial orientation. The direct electric field required to macroscopically change the orientation of polarization is similar in magnitude to the alternating field where the dynamic nonlinear dielectric permittivity reaches its maximum and polarization starts to saturate. This strongly suggests that the aforementioned maximum is related to the reorientation of polar nano entities and not to intrinsic tunability of dielectric permittivity by field.

The presented results question not only the commonly accepted narrative of a nonpolar ground state[1,2,20,60] but also the existence of an ergodic state[2,27] in PMN below $T_B$. We speculate that strain gradients[30], static polar nano entities[51], charge disorder[23], chemical ordered regions[61], or other mesoscopic inhomogeneities result in a separation of phase space into at least two parts, which individually act ergodically[62]. It cannot be ruled out that the plethora of sometimes contradicting experimental results reported in the literature is at least in part related to findings presented in this letter.

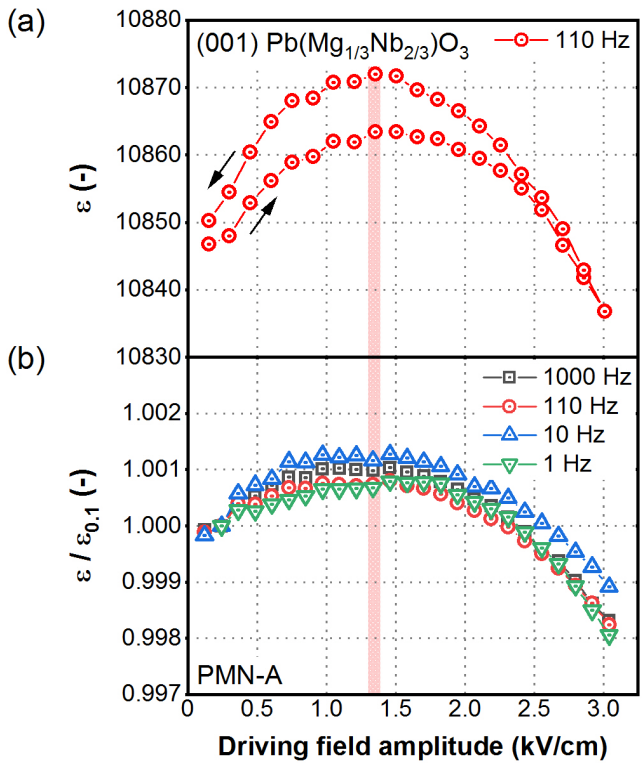

Figure 3: Dynamic dielectric permittivity ($\varepsilon$) of PMN-A as a function of driving field amplitude and frequency. Arrows indicate the direction of field cycling starting with increasing amplitude. The exact field amplitude value of maximum dielectric permittivity depends weakly on sample history and measurement conditions (Figure S4). It can be estimated to (1.50 ± 0.25) kV/cm with little frequency dependence between 1 Hz and 1 kHz. This value matches well with the experimentally determined field for polarization reversal $E_M$=(1.35 ± 0.05) kV/cm indicated as red

**hatched area. The coincidence between these two characteristic electric fields and the presence of dielectric hysteresis strongly suggest that the underlying mechanism is related to the switching of polar nano entities. Similar results were obtained for ceramic samples (Figure S3). (a) Very first measurement of field depended dynamic dielectric permittivity of crystal PMN-A. (b) Dynamic dielectric permittivity measurement of crystal PMN-A as a function of driving field frequency normalized to the first measurement point, i.e. the value at a driving field amplitude of ~0.1 kV/cm ($\varepsilon_{0.1}$).**

SUPPLEMENTARY MATERIAL

Additional pyroelectric and dielectric measurements are available in the supplementary material.

ACKNOWLEDGMENTS

This work was supported by the Swiss National Science Foundation (No. 200021_172525). We would like to thank Ms. Mizeret-Lad Jayshri for language revision of the manuscript.

AIP PUBLISHING DATA SHARING POLICY

The data that support the findings of this study are available within the article and its supplementary material. Raw data of this study are available from the corresponding author upon reasonable request.

**Supplementary Material for**
**"Macroscopic polarization in the ergodic phase of relaxor lead magnesium niobate"**
by Lukas M. Riemer, Kanghyun Chu, Yang Li, Hana Uršič, Andrew J. Bell, Brahim Dkhil, and Dragan Damjanovic

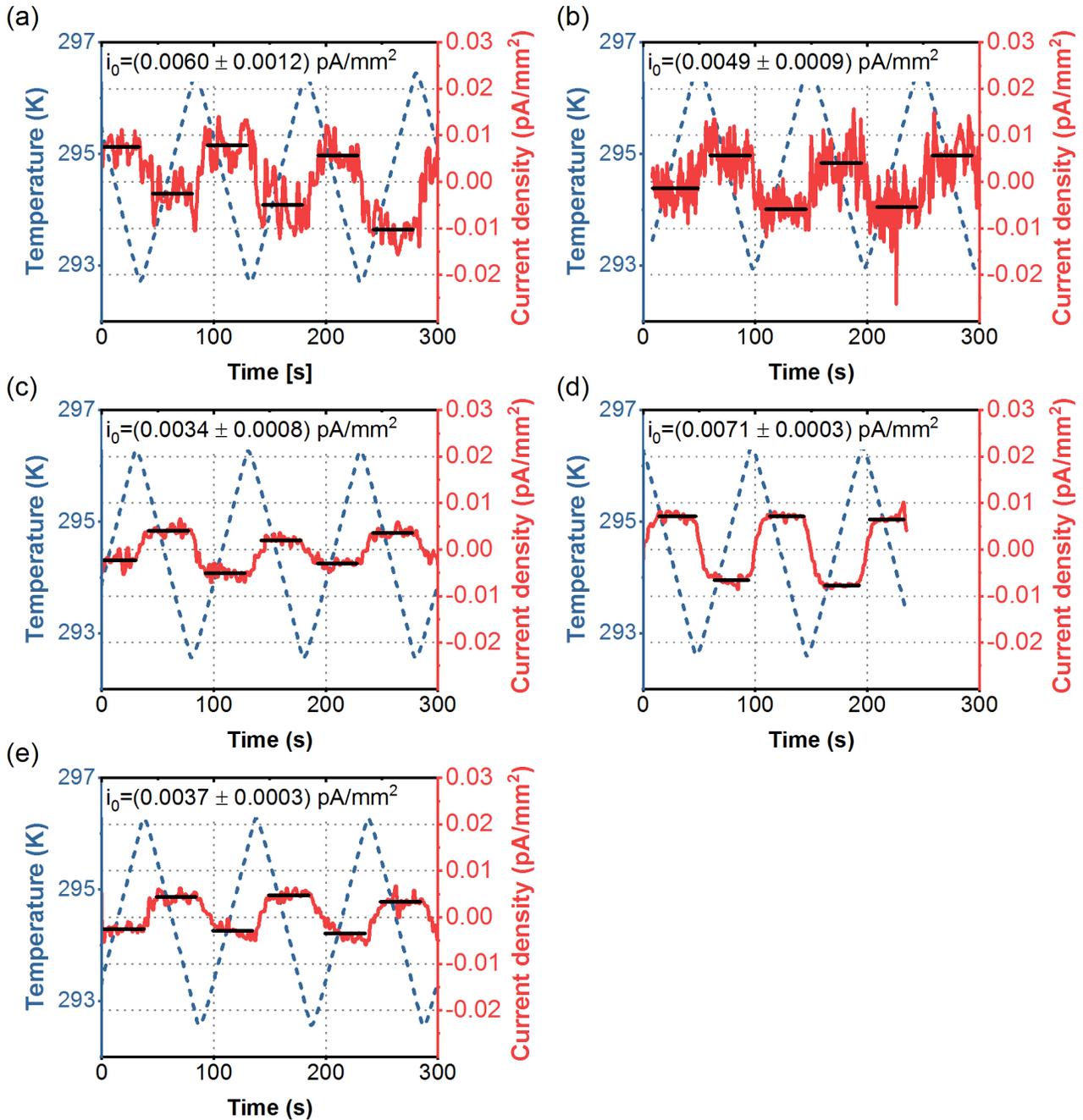

Figure S1: Pyroelectric response of Pb(Mg$_{1/3}$Nb$_{2/3}$)O$_3$ (PMN) crystals and ceramics. Current density amplitudes are fitted as indicated by black lines. The extracted averaged current density amplitudes $i_0$ are presented. Ceramics were annealed at 500 °C before measurements. (a) Pristine (001) PMN-A crystal, (b) pristine (001) PMN-B crystal, (c) annealed PMN ceramic (3.6 x 3.2 x 0.5) mm³, (d) annealed PMN ceramic (8.7 x 3.6 x 0.7) mm³, and (e) annealed PMN ceramic (3.6 x 3.5 x 1.0) mm³.

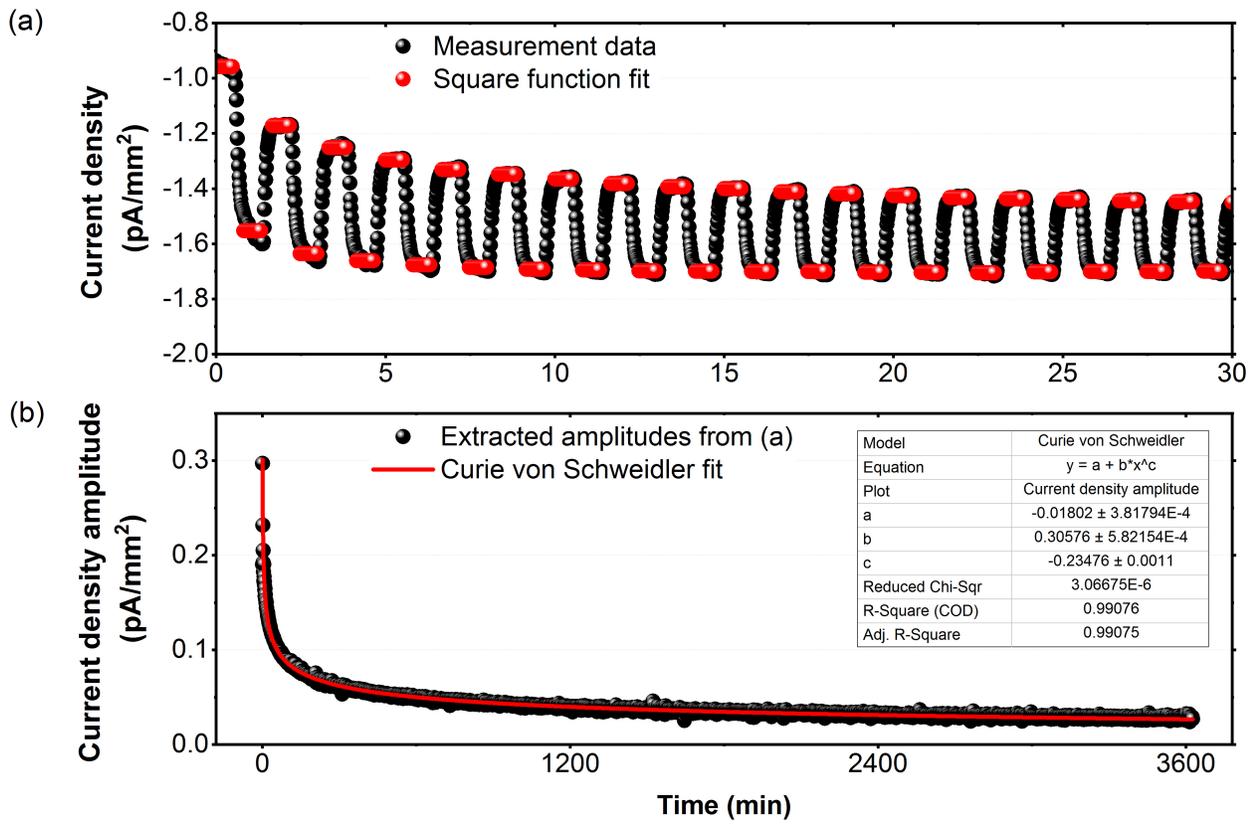

**Figure S2:** Decay of pyroelectric current response of (001) PMN-A crystal after poling at 10 kV/cm. (a) First 30 minutes after poling. The mean current $I_n$ within intervals of increasing and decreasing temperature are indicated in red. (b) Current density amplitudes $i_n$ are calculated as the absolute value of half the difference between adjacent current levels: $i_n = abs(I_n - I_{n+1})/2$. The temporal evolution of $i_n$ can be fit by the Curie von Schweidler law resulting in a power exponent n = -0.235 ± 0.001 in good agreement with results obtained from PFM measurements.[1]

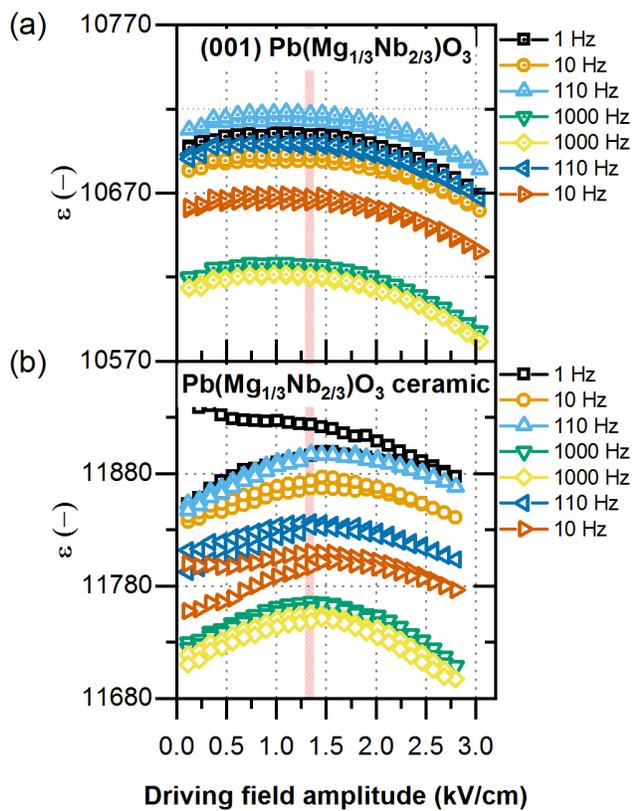

Figure S3: Sequence of field dependent dynamic dielectric permittivity ($\varepsilon$) measurements for a PMN crystal and ceramic performed in the order presented in the figure legend. The experimentally determined field for polarization reversal $E_M = (1.35 \pm 0.05)$ kV/cm is indicated as red hatched area. Maximum dielectric permittivity is obtained for a field amplitude of approximately $(1.50 \pm 0.25)$ kV/cm. (a) PMN-A crystal, and (b) ceramic with dimensions $(3.6 \times 3.3 \times 1.0)$ mm³.

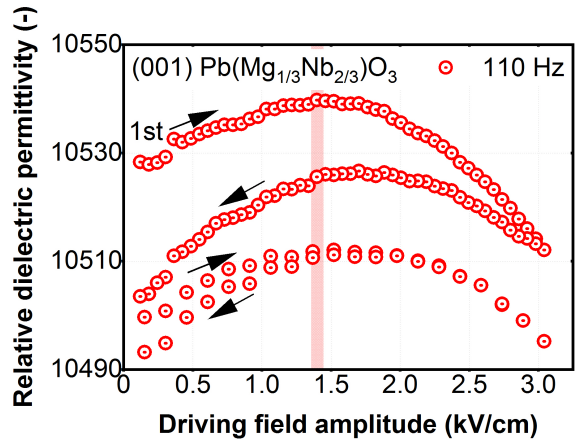

Figure S4: Example of hysteresis during field dependent dynamic dielectric permittivity (ε) measurements for crystal PMN-A. Arrows indicate the direction of field cycling starting with increasing amplitude. The experimentally determined field for polarization reversal $E_M=(1.35 \pm 0.05)$ kV/cm is indicated as red hatched area. Two subsequent measurements with different driving field amplitude increments are compared. During the first measurement the maximum in dynamic dielectric permittivity increases from ~1.4 kV/cm in the ascending branch to ~1.75 kV/cm in the descending branch. An increase of the field amplitude increment in the second measurement cycle significantly reduces dielectric hysteresis and confines it to field amplitudes below the maximum of permittivity. The previous increase of the field of maximum permittivity appears to be persistent.

SAMPLE PREPARATION

Crystal PMN-A was grown by a modified Bridgman approach. Precursor PMN ceramics were synthesized by two steps columbite precursor method using commercially available $(MgCO_3)_4 \bullet Mg(OH)_2 \bullet 5H_2O$, $Nb_2O_5$ and PbO powders. PMN pellets were sintered at 1423 K, charged into a cone-shaped platinum crucible and sealed with a platinum lid. The conical space of the platinum crucible was filled with a stoichiometric mixture of $MgNb_2O_6$ and $Pb_3O_4$ powders. A temperature gradient of 20 K/cm at the solid-liquid interface was obtained with a multi-zone tube furnace. Two upper zones of 1638 K and a lower zone of 1443 K were achieved at a heating rate of ~90 K/h. After the compounds were melted, the crucible was lowered down at a rate of 0.5 mm/h to crystalize PMN. After growth, the crystal was cooled down to ambient temperature with a cooling rate of 60 K/h. Crystal PMN-B was grown via flux method with $PbO-B_2O_3$ mixture as solvent. The mixed raw powders i.e. PbO, MgO, $Nb_2O_5$ and the solvent were placed in a Pt-Rh crucible before it was closed and heated to 1353 K. Then the crucible was cooled with ~0.5 K/h to 1173 K and with ~100 K/h to ambient temperature. The PMN crystal was then gently extracted from the crucible after etching in a hot acetic acid solution.

REFERENCES for Supplementary Material
[1] V.V. Shvartsman and A.L. Kholkin, Zeitschrift Für Kristallographie **226**, 108 (2011).